\documentstyle[epsfig,epsf,12pt]{article}
\def \bea{\begin{eqnarray}}
\def \beq{\begin{equation}}

\def \cm {{\cal M}}
\def \co {{\cal O}}
\def \eea{\end{eqnarray}}
\def \eeq{\end{equation}}
\def \ed{\epsilon_{\rm d}}
\def \eu{\epsilon_{\rm u}}

\def \s{\sqrt{2}}

\begin{document}
\begin{titlepage}

\large
\centerline {\bf Hierarchy and Anarchy in Quark Mass Matrices, or }
\centerline{ \bf Can Hierarchy Tolerate Anarchy?
\footnote{ Enrico Fermi Institute preprint EFI 01-26, hep-ph/0106335.
Submitted to Phys.\ Lett.\ B.}}
\normalsize
 
\vskip 2.0cm
\centerline {Rogerio Rosenfeld~\footnote{rosenfel@ift.unesp.br}}
\centerline {\it Instituto de Fisica Teorica - UNESP}
\centerline{\it Rua Pamplona, 145, 01405-900 Sao Paulo, SP, Brasil}
\smallskip
\centerline{and}
\smallskip
\centerline {Jonathan L. Rosner~\footnote{rosner@hep.uchicago.edu}}
\centerline {\it Enrico Fermi Institute and Department of Physics}
\centerline{\it University of Chicago, 5640 S. Ellis Avenue, Chicago, IL 60637}
\vskip 4.0cm
 
\centerline {\bf Abstract}
\vskip 1.0cm

The consequences of adding random perturbations (anarchy) to a baseline
hierarchical model of quark masses and mixings are explored.  Even small
perturbations of the order of $5\%$ of the smallest non-zero element can
already give deviations significantly affecting parameters of the
Cabibbo-Kobayashi-Maskawa (CKM) matrix, so any process generating
the anarchy should in general be limited to this order of magnitude.
The regularities of quark masses and mixings thus appear to be far from a
generic feature of randomness in the mass matrices, and more likely indicate
an underlying order.
\bigskip

\noindent
PACS Categories:  11.30.Hv, 12.15.Ff, 14.65.-q, 12.15.Hh

\vfill
\end{titlepage}

\newpage

\section{Introduction}

The origin of fermion masses and mixings is one of most important issues in
particle physics. Unfortunately, these parameters are inputs in the well-tested 
Standard Model. All one can do is to measure them as accurately as possible and
hope that in the future a more fundamental theory will actually be able to
predict their values. Many attempts have been made in this direction, with
grand unified theories (whether supersymmetric or not) being favorite candidates
\cite{models}.  However, such a fundamental theory could well be quite
complicated, with many new fields and couplings. Consequently, at low energies,
the observed fermion masses and mixings could be the result of a large number
of contributions.  If that were the case, one might expect that at low energy
scales the mass matrices of fermions would have a random nature.  This idea was
first suggested by Froggatt and Nielsen \cite{FNr}, who performed a statistical
study of fermion masses without success.

More recently, the idea of flavor anarchy was introduced in order to explain
new data on neutrino masses and mixings \cite{anarchy}, and several analyses
based on this idea have been performed \cite{rnd}. It was suggested that a
similar model could also explain masses and mixings in the quark and charged
lepton sectors. The purpose of this note is to study the robustness of such a
model for random quark matrices in the quark sector.  We take a baseline
model which approximately reproduces the observed masses and mixings and study
its sensitivity to random perturbations of the parameters.  In this way we
determine to what extent the observed regularities can survive effects which
may be purely coincidental or generic.  We find that quark masses and mixings
appear to be far from a generic feature of randomness in the mass matrices, and
more likely point to an underlying order.

\section{A simple ansatz}

The quark and charged lepton sectors are fundamentally different from the
neutrino sector because of the existence of a large mass hierarchy between
the families and the resulting small mixing.  We shall construct a ``baseline''
description of quark masses and mixings which incorporates several
approximate regularities.  For this purpose we begin with quark masses
evolved via the renormalization group to a common high mass scale $\mu =
M_Z$.  At this scale, the masses have been found to lie in the range \cite{FXr}
\begin{eqnarray}
m_{\rm u}(M_{\rm Z}) &=& 0.9 - 2.9    \;\; {\rm MeV}  \nonumber \\
m_{\rm c}(M_{\rm Z}) &=&  0.53 - 0.68 \;\; {\rm GeV}  \nonumber \\
m_{\rm t}(M_{\rm Z}) &=& 168-180 \;\; {\rm GeV}   \\
m_{\rm d}(M_{\rm Z}) &=& 1.8 - 5.3 \;\; {\rm MeV}  \nonumber \\
m_{\rm s}(M_{\rm Z}) &=& 35 - 100 \;\; {\rm MeV}  \nonumber \\
m_{\rm b}(M_{\rm Z}) &=& 2.8 - 3.0 \;\; {\rm GeV}        \nonumber
\end{eqnarray}
with corresponding ratios
$$
\sqrt{\frac{m_{\rm u}}{m_{\rm c}}} = \frac{1}{28} - \frac{1}{14} ~;~~
\sqrt{\frac{m_{\rm c}}{m_{\rm t}}} = \frac{1}{18} - \frac{1}{16}~;
$$
\beq
\sqrt{\frac{m_{\rm d}}{m_{\rm s}}} = \frac{1}{7.4} - \frac{1}{2.6} ~;~~
\sqrt{\frac{m_{\rm s}}{m_{\rm b}}} = \frac{1}{9.1} - \frac{1}{5.3}~.
\eeq
These masses thus are compatible with the hierarchy
\begin{equation} \label{eqn:qhier}
\frac{m_2}{m_3} = \frac{m_1}{m_2} = \epsilon^2 \; ,
\end{equation}
which we shall incorporate into our {\it ansatz} for quark mass matrices.
(This is certainly not a property of the charged leptons, for which \cite{GJ}
\beq \label{eqn:lhier}
\frac{m_2/3}{m_3} = \frac{m_1}{m_2/3} = \epsilon^2 \;
\eeq
is a better approximation.) For illustrative purposes we will adopt
$\epsilon_{\rm up} \equiv \eu = 0.07$ and $\epsilon_{\rm down} \equiv \ed
= 0.21$, which approximately reproduces the observed hierarchies.

We also seek a set of mass matrices reproducing the regularities
\beq
|V_{us}| \simeq |V_{cd}| \simeq {\cal O} \left( \sqrt{\frac{m_d}{m_s}} \right)
= \ed~~,~~~
|V_{cb}| \simeq |V_{ts}| \simeq {\cal O} \left( \frac{m_s}{m_b} \right)
= \ed^2~~~.
\eeq
The first of these was noted some time ago \cite{Gatto,Wbg,WZ}.  We further
wish to reproduce the hierarchy of the Cabibbo-Kobayashi-Maskawa (CKM)
matrix elements noted by Wolfenstein \cite{WP}, in which
\beq
|V_{us}| \simeq |V_{cd}| \simeq {\cal O} (\lambda)~,~~
|V_{cb}| \simeq |V_{ts}| \simeq {\cal O} (\lambda^2)~,~~
|V_{ub}| \simeq |V_{td}| \simeq {\cal O} (\lambda^3)~,
\eeq
($\lambda \simeq 0.22$), and with $|V_{ub}| < |V_{td}|$ as
favored by fits to data \cite{CKMfits}.

These regularities can be incorporated into a simple quark mass ansatz in a
basis which we call {\it hierarchical}:
\beq \label{eqn:mhier}
\cm_H \; = \; m_3 \left ( \matrix{
0	& \epsilon^3 e^{i \phi}	& 0 \cr
\epsilon^3 e^{-i \phi}	& \epsilon^2	& \epsilon^2 \cr
0	& \epsilon^2	& 1 } \right ) \; ,
\eeq
where $m_3$ denotes the mass eigenvalue of the third-family quark ($t$ or $b$).
Hierarchical descriptions of this type were first introduced by Froggatt and
Nielsen \cite{FN}.  The present ansatz is closely related to one described by
Fritzsch and Xing
\cite{FX,X}.  We shall not be concerned with relative coefficients of order 1
in different terms; for example, models in which the off-diagonal $\epsilon^2$
terms are multiplied by $\s$ may fit $|V_{cb}|$ somewhat better \cite{FX,RW}.
The $\epsilon^3$ terms in our model are
separate parameters in Ref.\ \cite{X}.  We have assumed the phase to be present
only in the $\epsilon^3$ terms; there is little sensitivity to phases in
the off-diagonal $\epsilon^2$ terms \cite{X}.  Our purpose is primarily to
construct an easily-manipulated ``cartoon'' version of the mass matrices, so
as to study the robustness of their predictions for masses and mixings under
random perturbations.

The eigenvalues of the matrix (\ref{eqn:mhier}) to order $\epsilon^4$ are given
by
\bea
\lambda_1 &= - m_3 \epsilon^4  \\             
\lambda_2 &= m_3 \epsilon^2        \\
\lambda_3 &= m_3 ( 1 + \epsilon^4 )
\eea
and are independent of the phase $\phi$. Therefore, in this ansatz the quark
masses naturally obey the hierarchy (\ref{eqn:qhier}).

The matrix (\ref{eqn:mhier}) can be made real by a unitary transformation 
$\cm_{\rm H}^{\rm R} = P \cm_{\rm H} P^{\dagger}$, where $P$ is given by
\begin{equation}  
P \; = \;  \left ( \matrix{
1	& 0	& 0 \cr
0	& e^{i \phi}	& 0 \cr
0	& 0	& e^{i \phi} } \right ) \; .
\end{equation}

The real symmetric matrix $\cm_{\rm H}^{\rm R}$ then can be diagonalized by an
orthogonal transformation $O^{T} \cm^{\rm R}_{\rm H} O$ where the matrix $O$ is
given to order $\epsilon^4$ (its columns are $u,c,t$) by:
\begin{equation}
O \; = \;  \left ( \matrix{ 1-\frac{\epsilon^2}{2} + \frac{3\epsilon^4}{8} &
 \epsilon - \frac{\epsilon^3}{2} & 0 \cr
- \epsilon + \frac{\epsilon^3}{2} & 1 - \frac{\epsilon^2}{2} -
 \frac{\epsilon^4}{8} & \epsilon^2 + \frac{\epsilon^4}{2} \cr
\epsilon^3 & - \epsilon^2 - \epsilon^4 & 1 - \frac{\epsilon^4}{2}} \right )
 \; .
\end{equation}

We will assume this same mass matrix ansatz for both the up and the down quark 
sectors. In this case, the Cabibbo-Kobayashi-Maskawa
(CKM) matrix is readily obtained by:
\begin{equation}
V_{\rm CKM}  = O^T_{\rm up} P_{\rm up} P^{\dagger}_{\rm down} O_{\rm down}~~.
\end{equation}

We first give approximate expressions for the CKM matrix elements:
\begin{eqnarray}
V_{\rm ud} & =&
1- (\epsilon_{\rm u}^2 +\epsilon_{\rm d}^2)/2 + e^{i \Delta} 
\epsilon_{\rm u} \epsilon_{\rm d} + \co(\epsilon^4) \\
V_{\rm us} &=&
\epsilon_{\rm d} - e^{i \Delta} \epsilon_{\rm u} + \co(\epsilon^3) \\
V_{\rm ub} &=&
\eu e^{i \Delta} (\eu^2 - \ed^2) + \co(\epsilon^5) \\
V_{\rm cd} &=&
 \epsilon_{\rm u} - e^{i \Delta} \epsilon_{\rm d} + \co(\epsilon^3) \\
V_{\rm cs} &=&
 e^{i \Delta}[1 - (\epsilon_{\rm d}^2 + \epsilon_{\rm u}^2)/2]
 + \epsilon_{\rm u} \epsilon_{\rm d} + \co(\epsilon^4) \\
V_{\rm cb} &=&
e^{i \Delta} ( \epsilon_{\rm d}^2 - \epsilon_{\rm u}^2) + \co(\epsilon^4) \\
V_{\rm td} &=&
\ed e^{i \Delta} (\ed^2 - \eu^2)  + \co(\epsilon^5) \\
V_{\rm ts} &=& 
 e^{i \Delta} ( \epsilon_{\rm u}^2 - \epsilon_{\rm d}^2) + \co(\epsilon^4) \\
V_{\rm tb} &=&
 e^{i \Delta} + \co(\epsilon^4) \; ,
\end{eqnarray}
where $\Delta \equiv \phi_{\rm u} - \phi_{\rm d}$.  These can be brought into
a form closer to the standard phase convention (see, e.g., \cite{WP}) by
multiplying the $c$ and $t$ rows by $e^{i(\chi - \Delta)}$ and the $s$ and $b$
columns by $e^{-i \chi}$, where $\chi = {\rm Arg}(\ed - e^{i \Delta}
\eu)$ is chosen so as to make $V_{us}$ and $V_{cd}$ real.  Then we
find, to leading order in small terms,
\begin{eqnarray}
V_{\rm ud} & =&
1- (\epsilon_{\rm u}^2 +\epsilon_{\rm d}^2)/2 + e^{i \Delta}
\epsilon_{\rm u} \epsilon_{\rm d} \\
V_{\rm us} &=&
|\epsilon_{\rm d} - e^{i \Delta} \epsilon_{\rm u}| \\
V_{\rm ub} &=&
\eu e^{i (\Delta - \chi)} (\eu^2 - \ed^2) \\
V_{\rm cd} &=&
- |\epsilon_{\rm d} - e^{i \Delta} \epsilon_{\rm u}| \\
V_{\rm cs} &=&
1 - (\eu^2 + \ed^2)/2 + e^{-i \Delta} \eu \ed \\
V_{\rm cb} &=&
\epsilon_{\rm d}^2 - \epsilon_{\rm u}^2 \\
V_{\rm td} &=&
\ed e^{i \chi}(\ed^2 - \eu^2) \\
V_{\rm ts} &=&
\eu^2 - \ed^2 \\
V_{\rm tb} &=& 1 \; .
\end{eqnarray}
The angles in the unitarity triangle can be expressed very simply in terms
of these quantities.  We find
$$
\alpha (= \phi_2) = \Delta~~,~~~
\beta (= \phi_1) = - \chi = {\rm tan}^{-1} \left( \frac{\sin \Delta}
{\ed/\eu - \cos \Delta} \right)~~,~~~
$$
\beq
\gamma (= \phi_3) = \pi - \alpha - \beta~~.
\eeq
These expressions also hold in more general versions of the present model
\cite{FX,X}.  Note that $|V_{ub}|$ and $|V_{td}|$ are specified entirely in
terms of the $\epsilon_{\rm u,d}$, with
\beq
|V_{cb}| = |V_{ts}| = \ed^2 - \eu^2 = 0.0392~~,~~~|V_{ub}/V_{td}| = \eu/\ed =
1/3~~~
\eeq
for our choice of parameters.  The shape of the unitarity triangle is
determined entirely by the magnitude of $V_{us}$, which changes as $\Delta$
varies.  It has been noticed previously that a value of $\Delta$ close to
$90^\circ$ gives a good fit to $|V_{us}|$ \cite{Wbg,WZ}.  For $\Delta = 
90^\circ$ we find $|V_{us}| = |V_{cd}| = (\ed^2 + \eu^2)^{1/2} = 0.221$.
The Wolfenstein parameters $\rho$ and $\eta$ defined by
\begin{eqnarray}
 \lambda \sqrt{\rho^2 +  \eta^2} &=  \frac{|V_{\rm ub}|}{|V_{\rm cb}|} 
&= \eu \\
\lambda \sqrt{(1/c^2-\rho)^2+\eta^2} &= \frac{|V_{\rm td}|}{|V_{\rm ts}|}
&= \ed \; ,
\end{eqnarray}
where $\lambda = |V_{\rm us}|$ and $c = 1 - \lambda^2/2 $, are given by:
\begin{eqnarray}
\rho &=& \frac{1}{2 c} \left( 1 - \frac{c^2}{\lambda^2} (\ed^2-\eu^2) \right)\\
\eta &=& \sqrt{\frac{\eu^2}{\lambda^2} - \rho^2 }\; .
\end{eqnarray}
With our choice of parameters, $\rho = 0.12$ and $\eta = 0.29$.

We now diagonalize the $u$ and $d$ quark mass matrices exactly (without the
above approximations), and calculate the resulting CKM matrix.
As an illustration, for the same parameters as above, we find
\begin{eqnarray}
|V_{\rm ud}| &=& |V_{\rm cs}| = 0.975 \\
|V_{\rm us}| &=& |V_{\rm cd}| = 0.222 \\
|V_{\rm cb}| &=& |V_{\rm ts}| = 0.0410 \\
|V_{\rm td}| = 0.0080~;~~|V_{\rm ub}| &=& 0.0029~;~~
|V_{\rm ub}/V_{\rm cb}| = 0.071 \\
\alpha = 98^\circ~;~~
\beta &=& 18^\circ~[\sin (2 \beta) = 0.58]~;~~\gamma = 64^\circ \\
\rho &=& 0.15~;~~ \eta = 0.29~~~.
\end{eqnarray} 

These predictions are reasonable and we will consider this simple model
as our baseline model.  We now perform random perturbations around it
(anarchy) to study the robustness of these predictions. 

\section{Anarchy}

The underlying theory of fermion masses is unlikely to guarantee that a 
particular mass matrix element is absolutely zero or has a fixed value given
by high energy symmetries of the theory. One could expect radiative effects or
small symmetry breaking parameters to contribute to a given mass matrix
texture. Since we don't have a complete theory to compute them, we will explore
the consequences of adding randomly small perturbations to our baseline model.
These perturbations are what we call anarchy.  We
will study just how much anarchy can our baseline model tolerate.

We will add to each element of the up and down mass matrices a random variable 
$\zeta$ with real part given by a random gaussian of zero average and a phase
given by a random variable uniformly distributed in the range $[0,2 \pi]$. We
ensure that the mass matrices remain hermitian by requiring that  $\zeta_{ij} =
\zeta^\ast_{ji}$.

The standard deviation of the real part of these random variables will be our
measure of anarchy. In fact, it would be natural to have the standard
deviations proportional to the smallest non-zero elements of the mass matrix,
namely $\sigma = \alpha \epsilon^3$, where $\sigma$
is the standard deviation of the gaussian distribution.

\begin{figure}
\centerline{\epsfxsize=0.75\hsize \epsffile{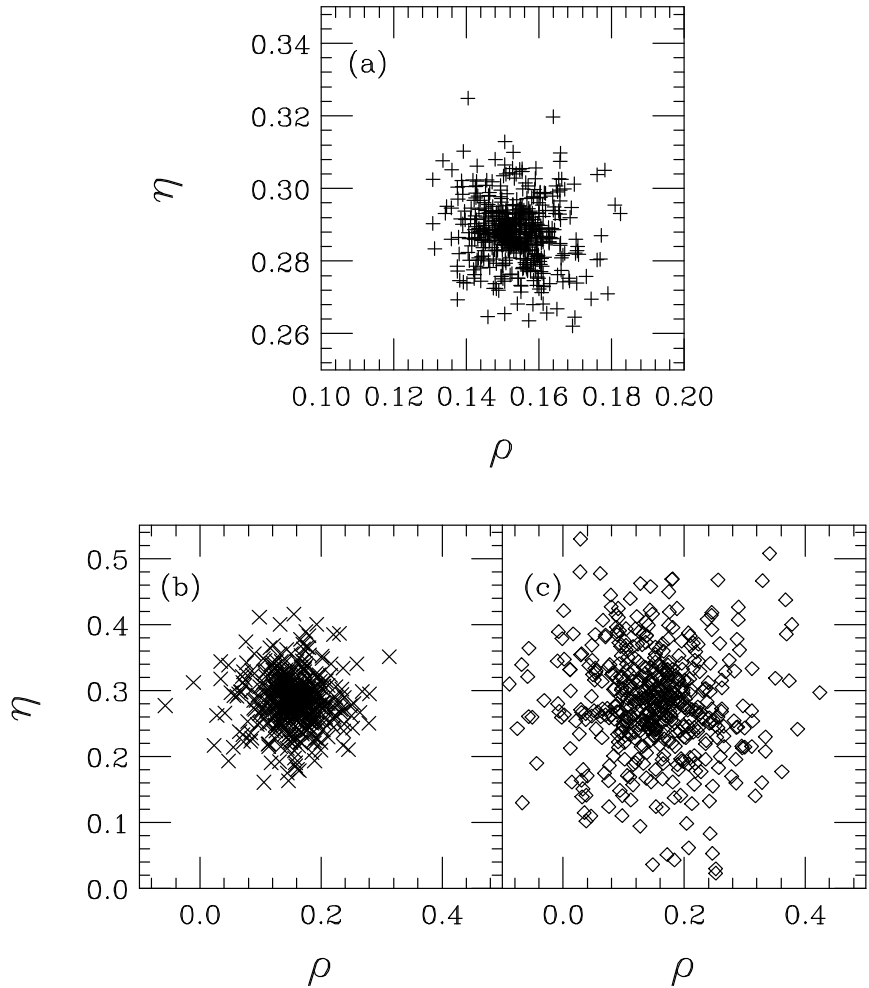}}
\caption{
\label{fig:re}Effect of anarchy in the $\rho - \eta$ plane for different values
of $\alpha$. (a) $\alpha = 0.01$; (b) $\alpha = 0.05$; (c) $\alpha = 0.1$.}
\label{re}
\end{figure}

\begin{figure}
\centerline{\epsfxsize=0.75\hsize \epsffile{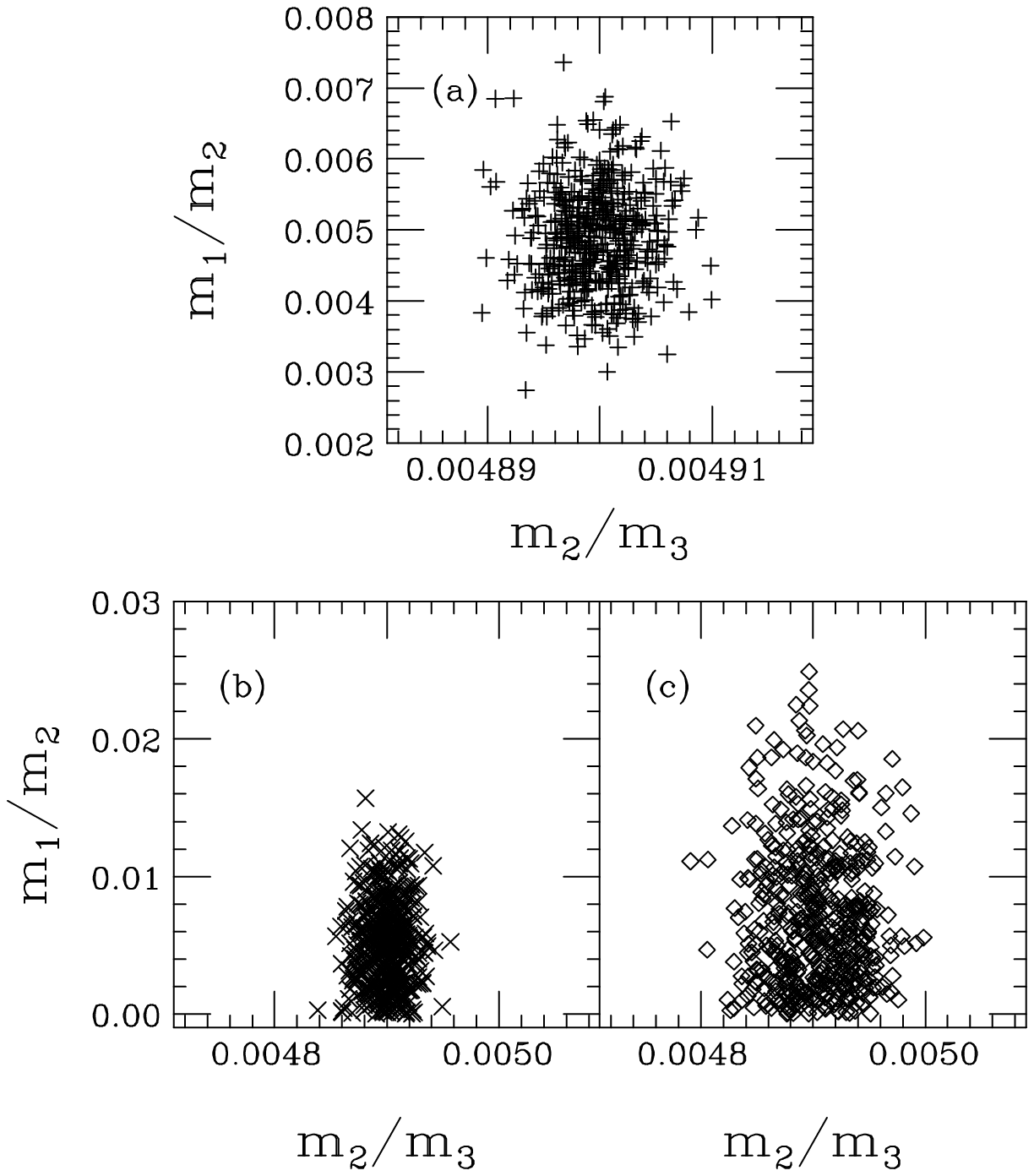}}
\caption{
\label{fig:mu}Effect of anarchy in the up sector in the 
$\frac{m_1}{m_2} - \frac{m_2}{m_3}$ 
plane for different values
of $\alpha$. (a) $\alpha = 0.01$; (b) $\alpha = 0.05$; (c) $\alpha = 0.1$.}
\label{mu}
\end{figure}

\begin{figure}
\centerline{\epsfxsize=0.75\hsize \epsffile{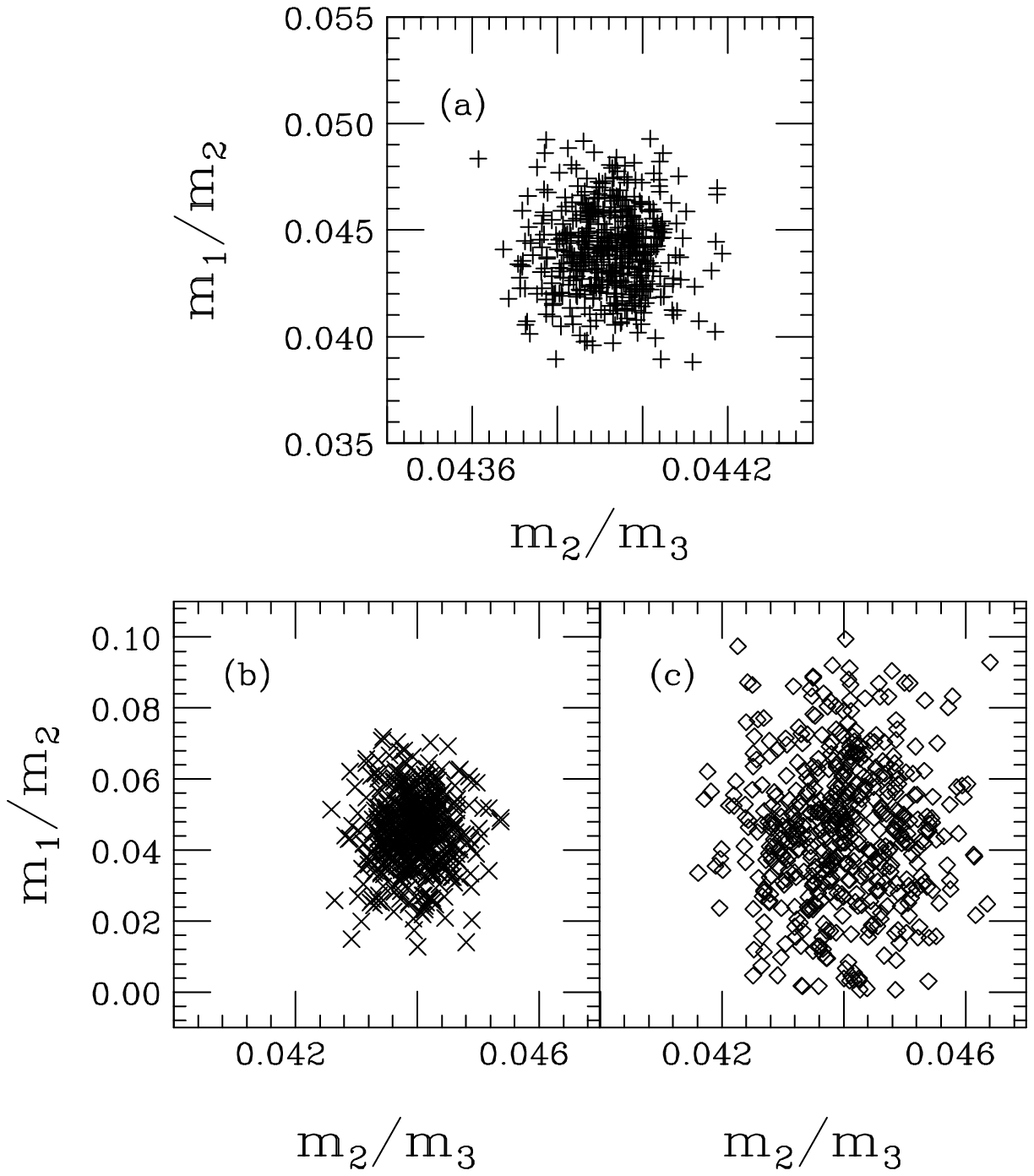}}
\caption{
\label{fig:md}Effect of anarchy in the down sector in the 
$\frac{m_1}{m_2} - \frac{m_2}{m_3}$ 
plane for different values
of $\alpha$.  (a) $\alpha = 0.01$; (b) $\alpha = 0.05$; (c) $\alpha = 0.1$.}
\label{md}
\end{figure}

In Figs.\ \ref{re}, \ref{mu}, and \ref{md}  we illustrate the effect of 
anarchy in the $\rho-\eta$ plane and the $\frac{m_2}{m_1} -
\frac{m_3}{m_2}$ planes for both up and down sectors for different values of
$\alpha$. As expected, the dispersion in the figures increases with $\alpha$.
The quantities $\rho$, $\eta$ and $\frac{m_3}{m_2}$ are more stable under
perturbations than the ratio of the two lightest quark masses,
$\frac{m_2}{m_1}$.

In order to measure the degree of variation of these parameters arising from
anarchy, we introduce the quantity $\kappa$, the ratio of the standard 
deviation $\sigma_X$ to the mean $\mu_X$ for a given parameter $X$ in a
simulation: $\kappa \equiv \sigma_X/\mu_X$.  In Table \ref{table} 
we show $\kappa$ for the different parameters generated by $1000$ simulations.

We can see that  $\kappa$ is roughly proportional to the parameter $\alpha$
determining the width of the gaussian distribution used to generate the
different perturbations. 
The sensitivity in the parameters increases in the following order:
$\left(\frac{m_2}{m_3}\right)_{\rm up}$, 
$\left(\frac{m_2}{m_3}\right)_{\rm down}$, $\eta$, 
$\left(\frac{m_1}{m_2}\right)_{\rm down}$, $\rho$ and
$\left(\frac{m_1}{m_2}\right)_{\rm up}$.
Not surprisingly, the most sensitive parameter is 
$\left(\frac{m_1}{m_2}\right)_{\rm up}$, 
since it involves the ratio of two small quantities.  Values of $\alpha < 0.1$
are required in order not to disturb the hierarchy significantly.  These
correspond roughly to ${\cal O}(\epsilon_d^5) \simeq {\cal O}(\lambda^5)$
entries in the mass matrices (\ref{eqn:mhier}).

\renewcommand{\arraystretch}{1.4}
\begin{table}
\caption{ Ratios $\kappa$ of the standard deviations to the
means for parameters generated by $1000$ simulations.}
\begin{center}
\begin{tabular}{c c c c} \hline \hline
          & $\alpha = 0.01$ &  $\alpha = 0.05$ &  $\alpha = 0.1$ \\ \hline
$\rho$  & 0.054        &   0.26        &   0.57         \\ 
$\eta$  & 0.029         &   0.14        &   0.28        \\
$\left(\frac{m_2}{m_3}\right)_{\rm up}$
& 0.00072         &   0.0034        &   0.0071        \\
$\left(\frac{m_1}{m_2}\right)_{\rm up}$
& 0.14        &   0.60        &   0.73        \\
$\left(\frac{m_2}{m_3}\right)_{\rm down}$
& 0.0021         &   0.010        &   0.021        \\
$\left(\frac{m_1}{m_2}\right)_{\rm down}$
& 0.046         &   0.24       &   0.45        \\ \hline \hline
\end{tabular}
\end{center}
\label{table}
\end{table}

\section{Conclusions}

We have performed a simple exercise of exploring the consequences of adding
random perturbations (anarchy) to a baseline hierarchical model of quark masses
and mixings. We find that even small perturbations, with gaussian distribution
of zero mean and standard deviations of the order of $5\%$ of the smallest 
non-zero element can already give deviations of $26\%$ in the $\rho$ parameter,
for instance.  Therefore, we conclude that any physics process generating
the anarchy, be it radiative corrections or small symmetry breaking parameters,
should be in general limited to this order of magnitude, unless some spurious
cancellations occur.  The regularities of quark masses and mixings thus
appear to be far from a generic feature of randomness in the mass matrices,
and probably indicative of an underlying order.

\section*{Acknowledgments}

We thank Hitoshi Murayama for constructive comments and Francesco Vissani for a
useful communication.
This work was supported in part by the United States Department of Energy
under Grant No.\ DE FG02 90ER49560. RR would like to thank FAPESP
(Funda\c{c}\~ao de Amparo \`a Pesquisa do Estado de S\~ao Paulo) for a grant to
visit the University of Chicago and thank also the Theory Group at the Physics
Department of the University of Chicago for the warm hospitality during his 
visit. 

\def \arnps#1#2#3{Ann.\ Rev.\ Nucl.\ Part.\ Sci.\ {\bf#1} (#3) #2}
\def \art{and references therein}
\def \cmts#1#2#3{Comments on Nucl.\ Part.\ Phys.\ {\bf#1} (#3) #2}
\def \cn{Collaboration}
\def \cp89{{\it CP Violation,} edited by C. Jarlskog (World Scientific,
Singapore, 1989)}
\def \econf#1#2#3{Electronic Conference Proceedings {\bf#1}, #2 (#3)}
\def \efi{Enrico Fermi Institute Report No.\ }
\def \epjc#1#2#3{Eur.\ Phys.\ J. C {\bf#1} (#3) #2}
\def \f79{{\it Proceedings of the 1979 International Symposium on Lepton and
Photon Interactions at High Energies,} Fermilab, August 23-29, 1979, ed. by
T. B. W. Kirk and H. D. I. Abarbanel (Fermi National Accelerator Laboratory,
Batavia, IL, 1979}
\def \hb87{{\it Proceeding of the 1987 International Symposium on Lepton and
Photon Interactions at High Energies,} Hamburg, 1987, ed. by W. Bartel
and R. R\"uckl (Nucl.\ Phys.\ B, Proc.\ Suppl., vol.\ 3) (North-Holland,
Amsterdam, 1988)}
\def \ib{{\it ibid.}~}
\def \ibj#1#2#3{~{\bf#1} (#3) #2}
\def \ichep72{{\it Proceedings of the XVI International Conference on High
Energy Physics}, Chicago and Batavia, Illinois, Sept. 6 -- 13, 1972,
edited by J. D. Jackson, A. Roberts, and R. Donaldson (Fermilab, Batavia,
IL, 1972)}
\def \ijmpa#1#2#3{Int.\ J.\ Mod.\ Phys.\ A {\bf#1} (#3) #2}
\def \ite{{\it et al.}}
\def \jhep#1#2#3{JHEP {\bf#1} (#3) #2}
\def \jpb#1#2#3{J.\ Phys.\ B {\bf#1} (#3) #2}
\def \jpg#1#2#3{J.\ Phys.\ G {\bf#1} (#3) #2}
\def \kaon{{\it Kaon Physics}, edited by J. L. Rosner and B. Winstein,
University of Chicago Press, 2000.}
\def \lg{{\it Proceedings of the XIXth International Symposium on
Lepton and Photon Interactions,} Stanford, California, August 9--14 1999,
edited by J. Jaros and M. Peskin (World Scientific, Singapore, 2000)}
\def \lkl87{{\it Selected Topics in Electroweak Interactions} (Proceedings of
the Second Lake Louise Institute on New Frontiers in Particle Physics, 15 --
21 February, 1987), edited by J. M. Cameron \ite~(World Scientific, Singapore,
1987)}
\def \kdvs#1#2#3{{Kong.\ Danske Vid.\ Selsk., Matt-fys.\ Medd.} {\bf #1},
No.\ #2 (#3)}
\def \ky85{{\it Proceedings of the International Symposium on Lepton and
Photon Interactions at High Energy,} Kyoto, Aug.~19-24, 1985, edited by M.
Konuma and K. Takahashi (Kyoto Univ., Kyoto, 1985)}
\def \mpla#1#2#3{Mod.\ Phys.\ Lett.\ A {\bf#1} (#3) #2}
\def \nat#1#2#3{Nature {\bf#1} (#3) #2}
\def \nc#1#2#3{Nuovo Cim.\ {\bf#1} (#3) #2}
\def \nima#1#2#3{Nucl.\ Instr.\ Meth. A {\bf#1} (#3) #2}
\def \npb#1#2#3{Nucl.\ Phys.\ B {\bf#1} (#3) #2}
\def \npps#1#2#3{Nucl.\ Phys.\ Proc.\ Suppl.\ {\bf#1} (#3) #2}
\def \npbps#1#2#3{Nucl.\ Phys.\ B Proc.\ Suppl.\ {\bf#1} (#3) #2}
\def \PDG{Particle Data Group, D. E. Groom \ite, \epjc{15}{1}{2000}}
\def \pisma#1#2#3#4{Pis'ma Zh.\ Eksp.\ Teor.\ Fiz.\ {\bf#1} (#3) #2 [JETP
Lett.\ {\bf#1} (#3) #4]}
\def \pl#1#2#3{Phys.\ Lett.\ {\bf#1} (#3) #2}
\def \pla#1#2#3{Phys.\ Lett.\ A {\bf#1} (#3) #2}
\def \plb#1#2#3{Phys.\ Lett.\ B {\bf#1} (#3) #2}
\def \pr#1#2#3{Phys.\ Rev.\ {\bf#1} (#3) #2}
\def \prc#1#2#3{Phys.\ Rev.\ C {\bf#1} (#3) #2}
\def \prd#1#2#3{Phys.\ Rev.\ D {\bf#1} (#3) #2}
\def \prl#1#2#3{Phys.\ Rev.\ Lett.\ {\bf#1} (#3) #2}
\def \prp#1#2#3{Phys.\ Rep.\ {\bf#1} (#3) #2}
\def \ptp#1#2#3{Prog.\ Theor.\ Phys.\ {\bf#1} (#3) #2}
\def \ppn#1#2#3{Prog.\ Part.\ Nucl.\ Phys.\ {\bf#1} (#3) #2}
\def \rmp#1#2#3{Rev.\ Mod.\ Phys.\ {\bf#1} (#3) #2}
\def \rp#1{~~~~~\ldots\ldots{\rm rp~}{#1}~~~~~}
\def \si90{25th International Conference on High Energy Physics, Singapore,
Aug. 2-8, 1990}
\def \zpc#1#2#3{Zeit.\ Phys.\ C {\bf#1} (#3) #2}
\def \zpd#1#2#3{Zeit.\ Phys.\ D {\bf#1} (#3) #2}

\end{document}